\documentstyle[12pt]{article}
\begin{document}
\baselineskip=22pt plus 0.2pt minus 0.2pt
\lineskip=22pt plus 0.2pt minus 0.2pt
\font\bigbf=cmbx10 scaled\magstep3
\begin{center}
 {\bigbf On the  metric operator for quantum cylindrical waves}\\

\vspace*{0.35in}

\large

Madhavan Varadarajan
\vspace*{0.25in}

\normalsize

{\sl Raman Research Institute,
Bangalore 560 080, India.}
\\
madhavan@rri.ernet.in\\
\vspace{.5in}
August 1999\\
\vspace{.5in}
ABSTRACT

\end{center}
Every (1 polarization) cylindrical wave  solution of 
vacuum general relativity
  is completely determined by a corresponding axisymmetric 
solution to the free
scalar wave equation on an auxilliary  2+1 dimensional flat spacetime. 
 The physical  metric at radius $R$ is determined  
by   the energy, $\gamma (R)$,  of the scalar
field in a box (in the flat spacetime) of radius $R$. 
In a recent work, among other important results, Ashtekar and Pierri have 
introduced a strategy to study the quantum geometry in this system, through 
a regularized quantum counterpart of $\gamma (R)$. 
We show that this 
regularized object is a densely defined 
symmetric operator, thereby correcting an error in their proof of this result. 
We argue that it admits a self adjoint extension and show that the operator,
unlike its classical counterpart, is not positive.

\pagebreak

\setcounter{page}{1}

\section*{1. Introduction} 

Quantization of cylindrically symmetric gravity
was initiated by Kucha{\v r}\cite{kuchar}.
The system has a `$z$'  (translational) and a `$\phi$' (rotational) Killing 
field.
{}From the general theory of 1 Killing reductions 
(see for eg. \cite{geroch,aamv}),
an analysis of the space of orbits of 
 the `$z$' directional Killing field yields an equivalent description
in terms of axisymmetric 2+1 gravity coupled to a 
massless axisymmetric scalar field. This latter
description has been studied by Allen \cite{allen} and recently, by 
Ashtekar and Pierri \cite{aamp}.

As a consequence of the axial symmetry, the scalar field satisfies the
free wave equation on an auxilliary {\em flat} 2+1 spacetime.
The physical
spacetime metric is then completely determined by the scalar field
and the quantization of the system is based on the 
{}Fock space associated with the free axially symmetric 
scalar field\cite{kuchar, allen,aamp}.

 Due to the axial  symmetry, the only dynamical 
3d metric components are the time-time and radial-radial ones,
namely $g_{TT}$ and  $g_{RR}$, where $T,R$ are the Einstein Rosen
 coordinates\cite{kuchar}.\footnote{Note that in \cite{kuchar}, 
$g_{TT},g_{RR}$ refer to the 4d metric
components in contrast to their 3d meaning here. The 4d and 3d components
are related by a multiplicative, scalar field dependent factor \cite{aamv}.}
   It turns out that 
$g_{RR} =-g_{TT}= e^{\gamma (R,T)}$, where $\gamma (R,T)$ is the energy 
of the scalar field in a disc of radius $R$ at time $T$. 

Among other issues explored and elucidated by Ashtekar and Pierri 
in \cite{aamp}, such as a careful treatement of the gauge fixing procedure, 
extraction of the true degrees of freedom and discussion of the issue of time 
and the Hamiltonian in quantum theory,  was the introduction of a strategy to
discuss quantum geometry. 



They found that since
$\gamma (R,T)$ is the energy of the field in a disc with a
{\em sharp} boundary at $R$, 
 it is neither a differentiable
function on the classical phase space nor is its quantum counterpart a
well-defined operator on the Fock space\cite{aamp}. They introduced a 
regulated version,
$\gamma (f_R)$, of $\gamma (R,T)$ which `softens' the sharp boundary by 
spatially
integrating the scalar field energy density against a smearing function, $f_R$,
which  is a smoothening of the step function
(the step function falls adbruptly to zero at $R$).

A key step in their discussion is the 
proof that the  regularized operator, $:{\hat\gamma (f_R)}:$,  is densely
defined. As we show in the appendix, 
it turns out that the sketch of the argument given in \cite{aamp}
to establish this result is incorrect.
In section 3, 
we rectify this situation by
providing a detailed proof of this result. We  also argue in favour of 
existence of self adjoint extensions of this (symmetric) operator. 

In section 4 we show that $:{\hat\gamma (f_R)}:$ admits negative 
expectation values. This is in contrast to its manifestly positive classical
counterpart, but not unexpected since it is well known (see eg. \cite{ford})
that stress energy quadratic forms do admit negative expectation values.

 Section 5 contains a discussion of our results and some comments.
 
Before we start on our results, we briefly summarize the 
relevant contents of \cite{kuchar,allen,aamp} in Section 2. 
We shall use units in which the gravitational constant in 2+1 dimensions, $G$,
Planck's constant, $\hbar$, and the speed of light, $c$, are unity.

\vspace{8mm}

\noindent {\bf Notation:} The quantum counterpart of a classical quantity,
$X$, is  denoted by $\hat{X}$. The time derivative of $X$ 
is denoted by $\dot{X}$,
its radial derivative by $X^\prime$ and its complex conjugate by
$X^*$. The adjoint of $\hat{X}$ is $\hat{X}^\dagger$ and its normal ordered
version is $:\hat{X}:$.

\section*{2. Review of classical and quantum cylindrical waves}
The cylindrical wave system is equivalent to 
axisymmetric 2+1 gravity coupled to a massless  axisymmetric
 scalar field. It is this equivalent description which is analysed in 
\cite{aamp}. We summarize the main results. 

The 2+1 dimensional  metric is 
\begin{equation}
ds^2 = e^{\gamma (R,T)}( -(dT)^2 + (dR)^2 ) + R^2 (d\phi)^2
\label{eq:cline}
\end{equation}
where $\phi$ is the  angular coordinate associated with the rotational
 Killing field,  $(R,T)$ are the Einstein Rosen 
coordinates \cite{kuchar} and 
\begin{equation}
\gamma (R,T) = {1\over 2} \int_{0}^{R}dr r
(\dot{\psi}(r,T)^2 + \psi^\prime(r,T)^2)
\label{eq:gammacl}
\end{equation}
where $\psi(R,T)$ is the axisymmetric solution to the flat 2+1 d 
wave equation,
\begin{equation}
-{\partial^2 \psi \over \partial T^2}
+{\partial^2 \psi \over \partial R^2}
+{1\over R}{\partial \psi \over \partial R} =0 .
\label{eq:waveeqn}
\end{equation}

The mode expansion of $\psi$ is 
\begin{equation}
\psi (R,T)= \int_0^{\infty}dk
                J_0(kR)(A(k)e^{-ikT} + A^*(k)e^{ikT})
\label{eq:classexp}
\end{equation}

The only nontrivial Poisson bracket between the mode coefficients is
\begin{equation}
\{ A(k),A^*(l)\}= -i\delta (l,k).
\end{equation}

Upon quantization, (\ref{eq:classexp}) goes over to  the
operator valued  distribution
\begin{equation}
\hat{\psi} (R,T)= \int_0^{\infty}dk
                J_0(kR)(\hat{A}(k)e^{-ikT} + \hat{A}^\dagger(k)e^{ikT})
\label{eq:quantexp}
\end{equation}
where $\hat{A}(k),\hat{A}^\dagger(k)$ are the annihilation and creation 
operators for mode $k$. They are represented in the standard way on the 
Fock space. The only nontrivial commutation relation between these operators
is
\begin{equation}
[\hat{A}(k),\hat{A}^{\dagger}(l)]= \delta(k,l).
\end{equation}

In this work, we shall concentrate on the behaviour of operators at $T=0$.
Let $\gamma (R):=\gamma (R,T=0)$. Then, from (\ref{eq:gammacl}) and
(\ref{eq:quantexp}), 
\begin{equation}
:\hat{\gamma} (R): = {1\over 2} \int_{0}^{\infty}dr r \theta(R-r)
:((\dot{\hat{\psi}}(r,0))^2 + (\hat{\psi}^\prime (r,0))^2): \; .
\label{eq:gammaq}
\end{equation}
where $\theta (x) =1$ if $x>0$ and vanishes elsewhere.
Direct computation \cite{aamp} shows that $||:\hat{\gamma} (R): |0>||$
diverges ($|0>$ is the Fock vacuum and `$||\;\;||$' 
denotes the Fock space norm) and so
 $:\hat{\gamma} (R): $ is not a well-defined operator on the Fock space.
The divergence comes about due to the discontinuity of the $\theta$ function.
The regulated version of this object, introduced in \cite{aamp} is
\begin{equation}
:\hat{\gamma} (f_R): = {1\over 2} \int_{0}^{\infty}dr r f_R(r)
:((\dot{\hat{\psi}}(r,0))^2 + (\hat{\psi}^\prime(r,0))^2)): \; .
\label{eq:gammareg}
\end{equation}
where 
$f_R(r)$ is a smooth function 
\footnote{For technical reasons, we require $f_R(r)$ to be a $C^{\infty}$
function. Note that the particular example of $f_R(r)$ in \cite{aamp} is
not $C^{\infty}$ at $r=R-\epsilon$ and is not suitable for our purposes.}
which equals 1 for $r\leq R-\epsilon$, smoothly decreases to zero for
$R-\epsilon < r < R+\epsilon$ and remains zero for $r\geq R+\epsilon$.
 $\epsilon$ is a parameter of dimensions of length. 

At T=0, the quantum line element becomes (see (\ref{eq:cline})) 
\begin{equation}
d\hat{s}^2 = e^{:\hat{\gamma} (f_R):}( -(dT)^2 + (dR)^2 ) + R^2 (d\phi )^2 .
\label{eq:qline}
\end{equation}
Substituting (\ref{eq:quantexp}) into (\ref{eq:gammareg}), we obtain   
\begin{eqnarray}
:\hat{\gamma} (f_R): &= & {1\over 2}\int_0^{\infty}dk_1\int_0^{\infty}dk_2
                       (2F_+(f_R,k_1,k_2) \hat{A}^{\dagger}(k_1)
                                          \hat{A}(k_2)\nonumber\\   
                    & +&
             F_-(f_R,k_1,k_2) (\hat{A}^{\dagger}(k_1)\hat{A}^{\dagger}(k_2)
                     +         \hat{A}(k_1) \hat{A}(k_2) )
\label{eq:gammaregq}
\end{eqnarray}
with 
\begin{equation}
F_{\pm}(f_R,k_1,k_2) =\pm k_1k_2\int_0^{\infty}dr r f_R(r)
                            (J_0(k_1r)J_0(k_2r)\pm J_1(k_1r)J_1(k_2r)).
\label{eq:Fpm}
\end{equation}.

The rest of this work is devoted to an 
 analysis of the properties of (\ref{eq:gammaregq}).

\section*{3. $:\hat{\gamma} (f_R):$ is densely defined}
In this section we show that $:\hat{\gamma} (f_R):$
is a (symmetric) densely defined operator. We first demonstrate that it maps
the vacuum into the Fock space. Next we show that it maps states of the form
$\int dk_1..dk_n g(k_1,..,k_n)\hat{A}^{\dagger}(k_1)..\hat{A}^{\dagger}(k_n)
|0>$, where $g(k_1..k_n)$ is a smooth function of rapid decrease in
$(k_1,..,k_n)$ space, into the Fock space. The above states, for all $n$,
along  with $|0>$
span a dense subspace of the Fock space. Finally, we show that the operator
is symmetric on this domain and argue for existence 
of its self adjoint extensions.

$:\hat{\gamma} (f_R): |0>$ is in the Fock space iff 
$||:\hat{\gamma} (f_R): |0>||$ is finite. But from (\ref{eq:gammaregq})
\begin{equation}
||:\hat{\gamma} (f_R): |0>||^2= 2\int dk_1dk_2|F_-(f_R, k_1,k_2)|^2 .
\end{equation}

We now demonstrate that $\int dk_1dk_2|F_-(f_R, k_1,k_2)|^2$ is finite. 
In fact, we shall prove the result for a general class of smearing functions,
$f(r)$, of which $f_R(r)$ is a particular case. Let $f(r)$ be $C^\infty$
on $(0,\infty)$ and let $f(r)$ and all its derivatives vanish faster than any 
power of $r^{-1}$ as $r\rightarrow \infty$. Further, let all derivatives of
$f(r)$ vanish faster than any power of $r$ as $r\rightarrow 0$. Denote the 
set of such functions 
by $S$.\footnote{Note that $f\in S$ is a rotationally symmetric
function of $(x,y)\in \bf {R^2}$ with  $x^2+y^2=r^2$. Its behaviour as 
$r\rightarrow 0$ ensures that it is a $C^\infty$ function on $\bf {R^2}$ at 
the origin. It is obviously $C^\infty$ elsewhere. Thus $f\in S$ is a smooth
function of rapid decrease on $\bf {R^2}$.}\\

\vspace{5mm}

\noindent {\em Lemma 1:} 
\begin{equation}
F_- := -k_1k_2\int_0^{\infty}dr r f(r)
                            (J_0(k_1r)J_0(k_2r)- J_1(k_1r)J_1(k_2r))
\label{eq:f-}
\end{equation}
has good enough ultraviolet behaviour in $k_1, k_2$ that 
\begin{equation}
\int_0^{\infty} dk_1\int_0^{\infty}
                      dk_2|F_-(f, k_1,k_2)|^2
\label{eq:finite}
\end{equation}
 is finite. \\
\vspace{3mm}

\noindent {\em Proof:} 
We shall use the Hankel transform,$G(k)$, of $rf(r)$ 
 with  respect to $J_1(kr)$. From the appendix, $G(k)\rightarrow 0$
faster than any power of $k$ as $k\rightarrow\infty$, where(see for 
eg. page 453 of \cite{watson})
\begin{eqnarray}
G(k) & = & \int_0^{\infty}r f(r)J_1(kr)r dr 
\label{eq:rf0}\\
rf(r)& = & \int_0^{\infty}G(k)J_1(kr)kdk .
\label{eq:rf}
\end{eqnarray}
Using (\ref{eq:rf}) in (\ref{eq:f-}), we obtain
\begin{equation}
-F_- = k_1k_2\int_0^{\infty}dkkG(k)\int_0^{\infty}dr J_1(kr)
                            (J_0(k_1r)J_0(k_2r)- J_1(k_1r)J_1(k_2r)).
\label{eq:f-1}
\end{equation}
From page 411 of \cite{watson},
\begin{eqnarray}
\int_0^{\infty}dr J_1(kr)J_0(k_1r)J_0(k_2r) 
                 &= {\theta\over k\pi} & \mbox{if $k_1,k_2,k$ form the sides
                              of a triangle, }
                                            \nonumber\\
              &  &  \mbox{with $\theta$ the angle between sides of} 
                                                           \nonumber\\ 
              &  &  \mbox{length $k_1$ and $k_2$} \nonumber \\
                 &= {1\over k}  &  \mbox{if $k>k_1 +k_2$} \nonumber\\
                 &= 0           &  \mbox{otherwise}   
\label{eq:j100}
\end{eqnarray}
and 
\begin{eqnarray}
\int_0^{\infty}dr J_1(kr)J_1(k_1r)J_1(k_2r) 
           &= {\sin\theta\over k\pi} & \mbox{if $k_1,k_2,k$ form the sides
                              of a triangle, }
                                            \nonumber\\
              &  &  \mbox{with $\theta$ the angle between sides of} 
                                                           \nonumber\\ 
              &  &  \mbox{length $k_1$ and $k_2$} \nonumber \\
                 &= 0           &  \mbox{otherwise}.   
\label{eq:j111}
\end{eqnarray}
Using (\ref{eq:j100}) and (\ref{eq:j111}) in (\ref{eq:f-1}),
we get
\begin{equation}
-F_- = k_1k_2\int_{|k_1-k_2|<k<k_1+k_2}dkG(k){(\theta -\sin\theta )\over \pi}
       + k_1k_2\int_{k>k_1+k_2}G(k)dk.
\end{equation}
We examine the behaviour of $F_-$ in the UV regime, i.e. 
as $k_1\rightarrow\infty$
or $k_2\rightarrow\infty$ or both $k_1,k_2\rightarrow\infty$.
From the Appendix A1, the last term gives rise to convergent contributions to 
(\ref{eq:finite}) because it falls off faster than any power of 
$k_1+k_2$. Hence, it suffices to examine the UV behaviour
of 
\begin{equation}
I:= k_1k_2\int_{|k_1-k_2|<k<k_1+k_2}dkG(k){(\theta -\sin\theta )\over \pi}
\end{equation}
We first consider the case $k_1>k_2, \;\; k_1\rightarrow\infty$.
$\Rightarrow k_1-k_2 <k < k_1+k_2$.
From the definition of $\theta$ (see (\ref{eq:j100})),
\begin{equation}
k^2= (k_1 -k_2)^2 +4k_1k_2 \sin^2 {\theta\over 2} \;\;\;\;\;\;\;\;
kdk=k_1k_2\sin\theta d\theta .
\label{eq:ktheta}
\end{equation}
\begin{equation}
\Rightarrow I= {(k_1k_2)^2\over \pi}\int_0^{\pi}d\theta
                 G(k)\sin\theta{(\theta -\sin\theta) \over k}.
\end{equation}
Let $I_1:={(k_1k_2)^2\over \pi}\int_{\theta_1}^{\pi}d\theta
                 G(k)\sin\theta{(\theta -\sin\theta) \over k}$ for some 
fixed small value of $\theta_1<<1$ independent of $k_1,k_2$.

For $\theta >\theta_1$ and large enough $k_1$, from (\ref{eq:ktheta})
\begin{eqnarray}
(i) & 0< k_2 < k_1^{1\over 3}  & 
          \Rightarrow (k_1-k_2)^2>k_1 \Rightarrow k>k_1^{1\over 2}
                                         \nonumber \\
(ii) &  k_1^{1\over 3}<k_2<k_1 & \Rightarrow k>2k_1^{2\over 3}
                                                 \sin{\theta_1\over 2}.
\end{eqnarray}
Thus for $\theta>\theta_1$, with fixed $\theta_1$ and large enough $k_1$,
$k>k_1^{1\over 2}$. Then from A1, $G(k)\rightarrow 0$ as 
$k_1\rightarrow\infty$ rapidly and so does $I_1$. Since $I_1$ gives convergent
contributions to (\ref{eq:finite}), it suffices to examine
$I_2:={(k_1k_2)^2\over \pi}\int_{0}^{\theta_1}d\theta
                 G(k)\sin\theta{(\theta -\sin\theta) \over k}$.

Let $p=k_1-k_2, \;\;dk_1dk_2=dk_1dp$ with $k_1\rightarrow\infty$. Then 
consider $I_2$ when \\
\noindent (a)$ k_1^{1\over 10} <p<k_1 $: 
\footnote{ The specific
choices of the values of the exponents as  ${1\over 10}$ here and of 
${4\over 9}$ in (\ref{eq:27}) are made only for concreteness.} 
From (\ref{eq:ktheta}), $k>p$.
Thus $k>k_1^{1/10}$ and from A1, $G(k)$ renders the contribution in this range 
to (\ref{eq:finite}) convergent. \\
\noindent (b)$0 < p< k_1^{1\over 10}$: 
$\Rightarrow k_1>k_2>k_1-k_1^{1\over 10}$ and $k_2\rightarrow \infty$. 
Put $I_2 = I_{2(i)} + I_{2(ii)}$ with 
\begin{equation}
I_{2(i)}:={(k_1k_2)^2\over \pi}\int_{(k_1k_2)^{-{4\over 9}}}^{\theta_1}d\theta
                 G(k)\sin\theta{(\theta -\sin\theta) \over k}.
\label{eq:27}
\end{equation}
For $(k_1k_2)^{-{4\over 9}}< \theta < \theta_1$ , there exists some $C>0$,
$C$ independent of $k_1, k_2$ such that for large enough $k_1$
\begin{equation}
k^2 > k_1k_2\sin^2{\theta\over 2}>Ck_1k_2 (k_1k_2)^{-{8\over 9}}.
\end{equation}
$\Rightarrow k>k_1^{1\over 3}$ and again , from A1, $G(k)$ renders this 
contribution convergent.

Finally consider 
\begin{equation}
I_{2(ii)}:={(k_1k_2)^2\over \pi}\int_0^{(k_1k_2)^{-{4\over 9}}}d\theta
                 G(k)\sin\theta{(\theta -\sin\theta) \over k}.
\end{equation}
Note that $|G(k)|$ is a bounded function.
$\Rightarrow$ There exists $C_1>0$ such that 
\begin{equation}
|I_{2(ii)}| <C_1(k_1k_2)^2\int_0^{(k_1k_2)^{-{4\over 9}}}d\theta
                                      {|\sin\theta(\theta-\sin\theta)|\over
                                      |\sqrt{k_1k_2}\theta |}
<C_1(k_1k_2)^2({\theta^4\over\sqrt{k_1k_2}})_{\theta=(k_1k_2)^{-{4\over 9}}}
\end{equation}
\begin{equation}
\Rightarrow |I_{2(ii)}| <C_1(k_1k_2)^{-{5\over 18}}
\end{equation}
\begin{equation}
\Rightarrow \int_0^{k_1^{1/10}}|I_{2(ii)}|^2dp <C_1^2\int_0^{k_1^{1/10}}
                dp (k_1(k_1-p))^{-{5\over 9}} <2 C_1^2k_1^{-{91\over 90}}
\end{equation}
\begin{equation}
\Rightarrow \int_{k_1\rightarrow\infty}dk_1\int_0^{k_1^{1/10}}|I_{2(ii)}|^2dp 
                                   <{2C_1^2\over k_1^{1\over 90}}\rightarrow 0.
\end{equation}
Thus $I_{2(ii)}$ also gives a convergent contribution.
Hence $\int_{k_2<k_1}|F_-|^2dk_1dk_2$ is finite. Since $F_-$ is symmetric 
under interchange of  $k_1$ and $k_2$, (\ref{eq:finite}) is finite.

We have also checked that no divergent terms appear in the expression for \\
\noindent $||:\hat{\gamma}(f):|g,n>|| \;\;$, 
\footnote{The notation is obvious - 
$f$ replaces $f_R$  in (\ref{eq:gammaregq})}
where 
\begin{equation}
|g,n>:=
\int dk_1..dk_n g(k_1,..,k_n)\hat{A}^{\dagger}(k_1)..\hat{A}^{\dagger}(k_n)
|0> , 
\end{equation}
with  $g(k_1,..,k_n)$ a smooth function of rapid decrease in $(k_1,..,k_n)$.

We sketch the main steps of the relevant calculation. The key point is that 
all terms in \\
\noindent $||:\hat{\gamma}(f):|g,n>|| \; \;$ involving the potentially 
divergent  expression 
$\int_0^{\infty} F_+(f,k_1,k_2)^2dk_1dk_2$ cancel. 
The remaining terms depend on $F_+$  through  either the expression
\begin{equation}
\int_0^{\infty} ds\; dp\; dq\;dk_2..dk_n F_+(f,p,s)F_+(f,s,q)
     g^*(p,k_2,..,k_n) g(q,k_2,..,k_n)
\end{equation}
or the expression
\begin{equation}
\int_0^{\infty} ds\; dt\; dp\; dq\; dk_3..dk_n F_+(f,s,t)F_+(f,p,q)
      g^*(s,p,k_3,..,k_n) g(t,q,k_3,..,k_n)
\end{equation}

Since $g(k_1..k_n)$ vanishes at 
infinity rapidly in each of its arguments, we can use Lemma 2 below to show 
that these terms are finite.

\noindent
$Lemma 2:$ Let 
$G_{\pm}(k_1):= 
\int_0^{\infty} dk_2F_{\pm} (f,k_1,k_2) g(k_2)$
 with
$g(k_2)$ a smooth function of rapid decrease as $k_2\rightarrow \infty$.
Then $G_{\pm}(k_1)$  falls off rapidly as $k_1\rightarrow \infty$.

\noindent $Proof:$ From (\ref{eq:rf0})- (\ref{eq:j111}),
\begin{equation}
\pm F_{\pm} = 
k_1k_2\int_{|k_1-k_2|<k<k_1+k_2}dkG(k){(\theta \pm \sin\theta )\over \pi}
       + k_1k_2\int_{k>k_1+k_2}G(k)dk.
\end{equation}
Let  $0<k_2< k_1-k_1^{1\over 10}, \;\; k_1\rightarrow \infty$.:\\
From (\ref{eq:ktheta}) and Appendix A1, the first term falls off rapidly
with large $k_1$. From A1, so does the second term. Hence 
\begin{equation}
G_{\pm}^{(1)}(k_1):= 
\int_0^{k_1-k_1^{1\over 10}} dk_2F_{\pm} (f,k_1,k_2) g(k_2)
\end{equation}
also vanishes rapidly as  $k_1\rightarrow \infty$.

Let $k_2> k_1-k_1^{1\over 10}, \;\; k_1,k_2\rightarrow \infty$.:\\
Since $J_0(x), J_1(x)$ are bounded functions of $x$, there exists $C_1$ such 
that
\begin{eqnarray}
|J_0(k_1r)J_0(k_2r)\pm J_1(k_1r)J_1(k_1r)|& < & C_1 \nonumber \\
\Rightarrow |F_{\pm} (f,k_1,k_2)| < C_f k_1 k_2,  & \;\;\;&
C_f := C_1 \int_{0}^{\infty} dr f(r) r .
\label{eq:Cf}
\end{eqnarray}
From(\ref{eq:Cf}) and the rapid fall off of $g(k_2)$ with large $k_2$, it is
clear that 
\begin{equation}
G_{\pm}^{(2)}(k_1):= 
\int_{k_1-k_1^{1\over 10}}^{\infty} dk_2F_{\pm} (f,k_1,k_2) g(k_2)
\end{equation}
falls off rapidly as $k_1\rightarrow \infty$.

Since $G_{\pm}(k_1)= G_{\pm}^{(1)}(k_1)+G_{\pm}^{(2)}(k_1)$, 
$G_{\pm}(k_1)$ also falls off rapidly as $k_1\rightarrow \infty$.
This completes the proof.

Using Lemmas 1 and 2, it can also be verified that all conceivable 
terms in \\
\noindent $||:\hat{\gamma}(f):|g,n>|| \; \;$ involving $F_-$ are also finite.

Thus, the result of these calculations is that $:\hat{\gamma}(f):$ 
maps the dense domain consisting of 
all $|g,n>$ and  $|0>$ into the Fock space.
Further, since $F_{\pm}(f,k_1,k_2)$ are real it is also symmetric and we have
shown that $:\hat{\gamma}(f):$
is a densely defined symmetric operator. 

We now argue in favour of existence of its self adjoint extensions along the  
lines of \cite{aamp}. We shall assume that there is no obstruction to an 
application of the standard treatments  \cite{glimmjaffe} of
free fields on ${\bf R^n}$, to the axisymmetric case. Although we do not
define the relevant function space below, we expect that a more careful 
treatement will convert our argument into a rigorous proof.

Let $\hat{C}$ be the complex conjugation 
operator in the Schroedinger representation on $L^2({\cal S}^{\prime},d\mu)$ 
where ${\cal S}^{\prime}$ is the  appropriate space of distributions
\footnote{The likely candidate for ${\cal S}^{\prime}$ 
is the topological dual to the space $S$ introduced earlier
in this section.} and  $d\mu$  the standard Gaussian measure associated with 
the restriction of the Laplacian in 2d to  rotational symmetry.
Then
$\hat{C}\hat{A}(k)=\hat{A}(k)\hat{C} $ 
and $\hat{C}\hat{A}^{\dagger}(k)=\hat{A}^{\dagger}(k)\hat{C} $. Also $\hat{C}$
leaves the domain of $:\hat{\gamma}(f):$ invariant.
This coupled with the reality of $F_{\pm}$ implies that $\hat{C}$ 
commutes with $:\hat{\gamma}(f):$. Then the theorem of Von Neumann 
\cite{reedsimon2}  cited  in \cite{aamp} shows that the operator admits self 
adjoint extensions.

\section*{4. Non-positivity of $:\hat{\gamma} (f_R):$}
We demonstrate the existence of states on which 
$:\hat{\gamma} (f_R):$ has negative expectation value.
The candidate states are motivated by those considered in \cite{ford}.
Let 
\begin{equation} 
|\phi> := {1\over 1+2\lambda^2}(|0>+ 
            \lambda\int_0^{\infty}
                 dp_1dp_2   g(p_1,p_2)A^{\dagger}(p_1)A^{\dagger}(p_1)
\end{equation}
with 
\begin{equation}
\int_0^{\infty}dp_1dp_2   g(p_1,p_2)g^*(p_1,p_2) =1 \;\;\;\;\;
   g(p_1,p_2)=g(p_2,p_1)
\end{equation}
Then a straightforward computation yields
\begin{eqnarray}
<\phi |:\hat{\gamma} (f_R):|\phi > &=& 
{4\lambda\over 1+2\lambda^2}
             \int_{0}^{\infty}dk_1   \int_{0}^{\infty} dk_2 \nonumber\\  
\big(2\lambda F_+(f_R, k_1, k_2) 
 \int_0^{\infty}dlg(l,k_2)g^*(k_1,l) 
            &+ & F_-(f_R, k_1, k_2)Re(g(k_1,k_2)\big)
\end{eqnarray}
where `$Re$' refers to `real part of'.

We shall choose $g(k_1,k_2)$ to be a step function. Thus, it is not 
contained in the dense domain defined in the previous section. However,
the choice of step function is just for pedagogy - as can easily be verified,
the conclusions are unaltered even if we replace the step function 
by a suitable smooth function of compact support which drops to zero
sufficiently fast.
Choose 
\begin{eqnarray}
g(k_1, k_2) &=& {1\over 2\delta} \;\;\; |k_1-k_0|<\delta,
                                       |k_2-k_0|<\delta \nonumber\\
            &=& 0 \;\;\;\; \mbox{otherwise}
\end{eqnarray}
where $k_0$ is a parameter which will be fixed later.
Then for sufficiently small $\delta$ to leading order in $\delta$,
\begin{equation}
<\phi |:\hat{\gamma} (f_R):|\phi > ={8\lambda\delta\over 1+2\lambda^2}
 (2\lambda F_+(f_R, k_0, k_0)
            +F_-(f_R, k_0, k_0))
\end{equation}
Now choose $k_0$ such that  $k_0(R+\epsilon)<<1$.
This enables us to use the small argument expansions of the Bessel functions 
appearing in (\ref{eq:Fpm}) and to leading order in $k_0(R+\epsilon)$
we obtain,
\begin{equation}
F_{\pm}(f_R, k_0,k_0) \sim \pm (k_0(R+\epsilon))^2 M, \;\;\;\;
M={1\over (R+\epsilon))^2} \int_0^{R+\epsilon}dr r f_R(r) .
\end{equation}
Thus, to leading order in $\delta$ and $k_0(R+\epsilon)$,
\begin{equation}
<\phi|:\hat{\gamma} (f_R):|\phi> ={8\lambda\delta 
                              (k_0(R+\epsilon))^2\over 1+2\lambda^2}
 (2\lambda -1)M .
\end{equation}
For $0<\lambda<{1\over 2}$, $:\hat{\gamma} (f_R):$ has negative expectation
values. Hence, in contrast to the positivity of its classical counterpart,
$:\hat{\gamma} (f_R):$ is {\em not} a positive operator.

\section*{5. Open issues and discussion}

By virtue of its solvability, the cylindrical wave midisuperspace is 
a very useful toy model for issues in full quantum gravity. In contrast to
the cosmological minisuperspace models, we can learn about the nonlinear 
{\em field} theoretic aspects of quantum gravity from this system.
To do this, it is important to understand the quantum metric operator.

We have shown that the regulated metric operator, 
$:\hat{\gamma} (f_R):$,
defined in \cite{aamp}
is indeed a densely defined symmetric operator. In contrast to its 
classical positivity,  $:\hat{\gamma} (f_R):$
is not a positive quantum operator.

Our results can also be viewed from the standpoint  of flat space 
quantum field theory. Indeed, apart from its significance for the quantum 
geometry of the cylindrical wave system, $:\hat{\gamma} (f_R):$ can 
 be thought of as a spatially smeared Hamiltonian operator for (axisymmetric)
free quantum field theory on a flat 2+1 spacetime. In {\em the absence  
of axisymmetry}, such spatially 
smeared Hamiltonians are well-defined quantum operators
in 1+1 dimensions but not in 2+1 (see appendix) 
or higher dimensions \cite{charlie,helfer}. Thus the {\em axisymmetric}
2+1 case is a `borderline' case in which all nonzero angular momentum 
modes of the field are switched off. Due to the absence of vacuum 
fluctuation contributions from  these modes, the smeared hamiltonian 
turns out to be a well-defined operator. Indeed, it is the presence of
these fluctuations which invalidates the idea of \cite{aamp} to
apply non axisymmetric 2+1 flat spacetime quantum field theory results
to the axisymmetric case.

The next step in understanding $:\hat{\gamma}(f_R):$, would be to see
if the operator is unbounded below or not. In this regard the results of 
Helfer \cite{helfer1} suggest that since $:\hat{\gamma}(f_R):$ is 
a well-defined operator, it should be bounded from below. If so, it would be 
of interest to calculate a lower bound for this operator, maybe by trying to 
generalise the techniques of \cite{ford1}.

In case there is a lower bound, it is possible to say more about the
properties of the operator. For example, by Theorem XII.5.2 of \cite{ds},
$:\hat{\gamma}(f_R):$
would then admit a self adjoint extension with the same lower bound.
Moreover, qualitative information regarding the spectrum may be obtained
using the min-max principle of \cite{reedsimon}.

Ultimately we would like to compare the spectrum of metric based operators 
(such as the 
determinant of the metric and the area operator) in 
this Fock representation of cylindrical waves with the spectrum of analogous
operators in the (non Fock) spin network representation of quantum gravity
\cite{arealee,areajurek}. In this regard note that 
for cylindrical waves, the (regulated)
area operator associated with an annulus between radii $R_1$ and $R_2$ is 
\begin{equation}
\hat{A} (R_1, R_2)= {1\over 2\pi}\int_{R_1}^{R_2}\sqrt{\hat {g}_{RR} }R dR
     ={1\over 2\pi} \int_{R_1}^{R_2}e^{:\hat{\gamma}(f_R):\over 2}RdR, 
\label{eq:area}
\end{equation}
where $\hat{g}_{RR}=e^{:\hat{\gamma}(f_R):}$.
The $\hat{g}_{RR}$ operators at different radii do not necessarily commute 
\cite{aamp} and it is difficult to infer anything about the spectrum of 
the area operator from that of $\hat{g}_{RR}$. 
 
We leave these and other issues (such as alternative regularizations 
of the metric/area operators) for possible future work.

\section*{Acknowledgments}

\noindent 
I gratefully acknowledge helpful discussions of
this material with A. Ashtekar, J. Samuel, S. Sinha,
 C. Torre and J. A. Zapata. I thank  Sukanya Sinha 
 for 
telling me about the results in \cite{ford}. I thank Professor Adimurti
and Professor Mythily Ramaswamy
for very useful discussions regarding the min-max principle.

\section*{Appendix}
\subsection*{A1}
Let $f(r) \in S$ (see beginning of section 3 for definition of $S$).
We show below that 
\begin{equation}
G(k)  =  \int_0^{\infty}r f(r)J_1(kr)r dr
\end{equation}
vanishes faster than any power of $k^{-1}$ as $k\rightarrow\infty$.

Note that (see footnote 3, section 3) as a rotationally symmetric function on 
$\bf R^2$, $f(r)$ is Schwartz. Hence its 2d Fourier transform, 
$F(k)$,
\footnote{ The Fourier transform of a rotationally symmetric function 
evaluated at the vector $\vec{k}$ depends only on the magnitude `$k$' of
$\vec{k}$ as can be seen from (\ref{eq:50}).}
 is also 
Schwartz (see for eg. \cite{reedsimon3}).
\begin{equation}
F(k) ={1\over 2\pi}\int_0^{\infty}drr\int_0^{2\pi}d\theta
                     f(r)e^{ikr\cos\theta}.
\label{eq:50}
\end{equation}
From \cite{sommerfeld}
$J_0(kr) ={1\over 2\pi}\int_0^{2\pi}d\theta
                     e^{ikr\cos\theta}$.Hence 
\begin{equation}
F(k)= \int_0^{\infty}drrf(r)J_0(kr)  .
\end{equation}
Thus, the Hankel transform of $f(r)$ with respect to $J_0(kr)$
is the same as its 2d Fourier transform. Hence the Hankel transform $F(k)$
is a function of rapid decrease.
Using properties of the Hankel transform and the Bessel functions\cite{watson}
we get
\begin{equation}
f(r)=\int_0^{\infty}dk k J_0(kr)F(k) ={1\over r}\int_0^{\infty}
                                     dkF(k){d\over dk}(J_1(kr)k)
\end{equation}
A by parts integration using the asymptotic properties of $F(k)$ gives
\begin{equation}
\Rightarrow rf(r)=-\int_{0}^{\infty}dk k J_1(kr){d\over dk} F(k)
\end{equation}
But $G(k)=- {d\over dk} F(k)$ is a function of rapid decrease
because $F(k)$ is a function of rapid decrease.

Q. E. D.

\subsection*{A2}
In this appendix we identify the error in the proof of \cite{aamp},
 which attempts to show that $||:\hat{\gamma}(f_R):|0>||$ is finite.
\footnote{
We use unpublished  results of C. Torre below and we thank him for allowing us
to reproduce his results here.}
We assume familiarity with the relevant part of \cite{aamp}.

The claim in \cite{aamp} is that 
\begin{equation}
I:= \int d^2\vec{k_1}d^2\vec{k_2}|G_-(f_R, \vec{k_1},\vec{k_2})|^2
\label{eq:I}
\end{equation}
is finite with 
\begin{equation}
G_-(f_R, \vec{k_1},\vec{k_2}) = - {(k_1k_2 +\vec{k_1}\cdot\vec{k_2})
                           \over \sqrt{k_1k_2}}f(\vec{k_1}+\vec{k_2})
\end{equation}
where $f(\vec{X})$ is the 2d Fourier transform of $f_R(r)$ evaluated at 
wave vector $\vec{X}$ and the magnitude of $\vec{X}$ is denoted by 
$|\vec{X}| = X$.  We shall show that, in fact, $I$ diverges.

Note that (\ref{eq:I}) converges iff it converges absolutely. Therefore,
in what follows, 
we are permitted to change
 variables and 
orders of integration \cite{rs1.21}.

Put $\vec{k_1}+\vec{k_2}= \vec{q}$. Let $\theta$ be the angle between 
$\vec{k_1}$ and $\vec{q}$  and let $\theta_1$ be the angular 
coordinate of $\vec{k_1}$. Then, with $0<\theta_1<2\pi$ and 
$\theta_1 <\theta <\theta_1+2\pi$,
\begin{equation}
d^2\vec{k_1}d^2\vec{k_2} = k_1 dk_1 d\theta_1 qdqd\theta .
\label{eq:c1}
\end{equation}
Then,
\begin{equation}
|G_-(f_R, \vec{k_1},\vec{k_2})|^2= |f(\vec{q})|^2
                                   |\Phi_{\vec{q}}(\vec{k_1})|^2
\label{eq:c2}
\end{equation}
with 
\begin{equation}
\Phi_{\vec{q}}(\vec{k_1}) =
{(k_1|\vec{q}-\vec{k_1}|
 +\vec{k_1}\cdot(\vec{q}-\vec{k_1}))
                           \over \sqrt{{k_1}|\vec{q}-\vec{k_1}|}}.
\end{equation}
We now expand the numerator of 
$\Phi_{\vec{q}}({\vec{k_1}})$ 
in powers of
${q\over k_1}$ to get the leading behavior as $q/k_1\rightarrow 0$:
\begin{equation}
\Phi_{\vec{q}}({\vec{k_1}}) \sim {q^2\sin^2\theta \over 2 k_1} .
\label{eq:c3}
\end{equation}
From (\ref{eq:c1}), (\ref{eq:c2}) and (\ref{eq:c3}), $I$ diverges 
logarithmically
with $k_1$ as $k_1\rightarrow\infty$. (Note that when $R=\infty$, $f(\vec{q})$
has support only at $q=0$ and hence the integral vanishes. This is as 
expected because in this limit $f_{R=\infty}=1$ and 
$:\hat{\gamma}(f_\infty):$  is just the total energy operator for the scalar
field. But for any finite $R$, the integral $I$ diverges.)

\end{document}